\documentclass[aps,preprint]{revtex4}
\usepackage{epsfig,amssymb,colordvi}

\begin{document}
\title[]{Doping dependence of phase coherence between superconducting Bi$_2$Sr$_2$CaCu$_2$O$_{8+\delta}$ grains}

\author{G. C. \surname{Kim}}
\author{M. \surname{Cheon}}
\author{Y. C. \surname{Kim}}
\email{yckim@pusan.ac.kr}
\affiliation{Department of Physics, Pusan National University, Busan 609-735, Korea}

\received{}

\begin{abstract}
In the present work, we report the new findings on the doping level dependence of the phase coherence between superconducting Bi$_2$Sr$_2$CaCu$_2$O$_{8+\delta}$ (Bi-2212) grains. The experimental results from the strongly underdoped and overdoped regimes deviated from the expectation based on the doping level dependence of the superfluid density at $T$ = 0 K. These findings appear to be governed by interplay between competing orders inside the superconducting dome of cuprate superconductors. Two quantum critical points are likely to exist at the underdoped and overdoped regimes beneath the superconducting dome.  \\
\\


\end{abstract}

\maketitle

A quantum phase transition (QPT) is a phase transition driven by quantum fluctuations at 0 K without thermal fluctuations \cite{R1}. The QPT between two or more phases can be tuned by varying the external parameters such as pressure, strength of the magnetic field, or the density of electrons. As the external parameter is varied at 0 K, QPT occurs at some critical point, which is called the quantum critical point (QCP). Although the QCP exist only at 0 K, the influence of quantum criticality extends over a broad regime, $i.e.$ a quantum critical region, above 0 K in the phase diagram. Close to the QCP, the correlation length diverges with a power-law dependence on the distance from the QCP \cite{R1}. 

Sebastian $et$ $al.$ \cite{R2} reported that the extrapolations of $T_{MI}$, at which the in-plane resistivity reaches its lowest value before diverging logarithmically at low temperatures, and the Fermi temperature, $T_F$, which was extracted from the quantum oscillation experiments, to 0 K for the underdoped YBa$_2$Cu$_3$O$_{6+\delta}$ (YBCO) samples, revealed an intercept at $p$ $\sim$ 0.08 below the superconducting dome, where $p$ is the hole doping level. Vishik $et$ $al.$ \cite{R3} showed from angle-resolved photoemission spectroscopy (ARPES) of Bi-2212 that for 0.076 $<$ $p$ $<$ 0.19, gaps at the near-nodal and intermediate momenta were independent of doping, but for $p$ $<$ 0.076, the Fermi surface was fully gapped and the gap anisotropy decreases. Grissonnanche $et$ $al.$ \cite{R4} reported that the upper critical field at $T$ = 0 K, $H_{c2}(0)$, versus the $p$ phase diagram for YBCO consists of two peaks, which are located at $p_{1}$ $\sim$ 0.08 and $p_{2}$ $\sim$ 0.18. These results strongly indicate the presence of two QCPs at the underdoped and overdoped regimes in the phase diagram of the cuprate superconductors.    

Motivated by these findings, this study investigates the doping dependence of phase coherence between the superconducting Bi-2212 grains over a sufficiently wide range of $p$. As $p$ decreases from the optimal doping to the underdoped limit of the superconducting dome, the superfluid density at $T$ = 0 K decreased, whereas the phase coherence between the underdoped superconducting Bi-2212 grains was enhanced. In the overdoped regime, a sudden reversal in the doping dependence of phase coherence between the superconducting Bi-2212 grains was observed. These results strongly support the presence of QCPs at $p_
{c1}$ $\simeq$ 0.075 and $p_{c2}$ $\simeq$ 0.190 in the superconducting dome of the cuprate superconductors. 

Granular samples of Bi-2212 with different $p$ values were prepared using the solid state reaction method, as described elsewhere \cite{R5}. Each of the samples was sintered twice with intermediate grinding to enhance the carrier homogeneity. The doping level for each sample was changed by partially substituting Ca with Y. The resistivity of the samples was measured with a current of $I$ = 1.0 mA and the ac susceptibilities were measured under a small field of $H_{rms}$ = 0.1 Oe to exclude the effects of the magnetic field as much as possible. The superconducting transition temperature, $T_c$, was considered to be the temperature at which the diamagnetic signal appears. The $p$ values for all the samples were calculated from the empirical relationship between $T_c$ and $p$, $T_{c}/T_{c,max} = 1 - 82.6(p - 0.16)^2$, where $T_{c,max}$ = 94 K for Bi-2212 \cite{R6}.

In a granular superconductor, as the phases of the different superconducting grains are coupled by the Josephson tunneling of Cooper pairs, the resistivity approaches zero \cite{R7}. Therefore, the temperature, at which the resistivity disappears, was used as a measure of the phase coherence between the superconducting grains near $T_{c}$. The temperature is denoted as $T_{\phi}$, which is the same as or lower than $T_c$, depending on the material. For granular YBCO and La$_{2-x}$Sr$_x$CuO$_4$ (LSCO) with a low anisotropy ratio, $T_{\phi}$ is almost equal to $T_{c}$, which means that the long-range phase coherence between the superconducting grains is established with a superconducting transition almost simultaneously. On the other hand, for Bi- and Tl-based cuprate superconductors close to two dimensional (2D) systems due to the extremely anisotropic property, $T_{\phi}$ is much lower than $T_{c}$ due to the enhanced thermal fluctuations. Moreover, as $p$ decreases from the overdoped to underdoped regimes, the 2D anisotropic nature in Bi- and Tl-based cuprate superconductors become more remarkable, which makes the phase more sensitive to the thermal fluctuations. Therefore, Bi-2212 is a good candidate for studies of the relationship between $T_{\phi}$ and $T_{c}$ as a function of $p$.

The superfluid density at $T$ = 0 K, $\rho_s(0)$, is a measure of the phase stiffness of the superconducting order parameter for the thermal fluctuations, which destroys the long-range order \cite{R8}. This $\rho_s$ is closely related to the penetration depth, $\lambda$, as $\rho_s(T)$ $\equiv$ $\lambda^{-2}(T)$. The values of $\lambda$ for strongly underdoped Bi-2212 were calculated from the reversible magnetization measurements, as described elsewhere \cite{R9}. As shown in the inset in Fig. 1, as $p$ decreases to the lower limit of the superconducting dome, $\lambda^{-2}(0)$ , $i.$$e.$, $\rho_s(0)$, approaches zero, which is consistent with the results reported by Anukool $et$ $al.$ \cite{R10} for Bi-2212. This suggests that the temperature range of the phase incoherence between the superconducting grains increases with decreasing $p$ due to the enhanced phase fluctuations.

Figure 1 shows the temperature dependence of the resistivity and ac susceptibility for Bi-2212 with different $p$ values ranging from $p$ = 0.068 to 0.213. Despite the significant data on the samples, this paper presents the data from representative samples for the sake of clarity. In contrast to the expectation of enhanced phase fluctuations based on the decrease in $\rho_s(0)$ in the underdoped regime mentioned above, Fig. 1 shows two important features from the temperature dependence of the ac susceptibility and resistivity for each sample. First, for $T$ $<$ $T_{\phi}$, while the diamagnetic signals of the overdoped and moderately underdoped Bi-2212 increase slowly and nonlinearly, those for strongly underdoped Bi-2212 increase rapidly and linearly, similar to a single crystal with a single superconducting phase. Second, as $p$ goes from the overdoped to the strongly underdoped regime, the difference between $T_c$ and $T_{\phi}$, $ \Delta T$ $\equiv$ $T_c-T_{\phi}$, decreases, reaching almost zero at $p$ $\simeq$ 0.073, and then increasing again. Therefore, these results suggest that $\rho_s(0)$ is not the only factor determining the phase coherence between the strongly underdoped superconducting Bi-2212 grains.

To assess the degree of phase coherence between the superconducting grains near $T_{c}$, the $p$ dependence of $ \Delta T$ normalized by $T_c$, $\Delta T/T_c$, is plotted in Fig. 2. The results show that the magnitude of $\Delta T/T_c$ remains  almost constant at 0.140 near $p$ = 0.15, and decreases by approximately one order of magnitude at $p$ $\simeq$ 0.073, even though $\rho_s(0)$ is reduced significantly, and then increases to 0.102 again at $p$ $\simeq$ 0.068. Therefore, as soon as the superconductivity (SC) emerges for Bi-2212 at $p$ $\simeq$ 0.075, the long-range phase coherence between the superconducting grains is established, as in conventional superconductors. On the other hand, in the overdoped regime ($p$ $>$ 0.16, at which $T_c$ reaches its maximum), the magnitude of $\Delta T/T_c$ shows a non-monotonic $p$ dependence with a broad dip with a minimum, rather than zero, at $p$ $\simeq$ 0.19. 

In general, $\rho_s(0)$ is independent of $T_c$, because $T_c$ ($\rho_s(0)$) is determined by the maximum superconducting gap (density of state) on  the Fermi surface \cite{R11}. However, a linear relationship is generally observed between $T_c$ and $\rho_s(0)$ in the moderately underdoped regime \cite{R12}. Therefore, $T_c$/$\rho_s(0)$ on Bi-2212 is superimposed as a function of $p$ in Fig. 2 using $T_c$ and $\rho_s(0)$ given in Ref. 13 and in the inset of Fig. 1, where, for comparison, a constant is multiplied such that values of $T_c$/$\rho_s(0)$ near $p$ = 0.15 is approximately 0.140 K$\cdot$$\mu$m$^2$. This $T_c$/$\rho_s(0)$ is used as a measure of the strength of the fluctuations \cite{R11}. Fig. 2 shows good agreement between the $p$ dependence of $\Delta T/T_c$ and that of $T_c$/$\rho_s(0)$ from $p$ $\approx$ 0.10 to 0.19, indicating that the same mechanism for the $p$ dependence of $\Delta T/T_c$ is at play in 0.10 $<$ $p$ $<$ 0.19. On the other hand, $\Delta T/T_c$ deviates downward from the $p$ dependence of $T_c$/$\rho_s(0)$ at $p$ $\simeq$ 0.10 in the underdoped regime and upward at $p$ $\simeq$ 0.19 in the overdoped regime, . 

Because SC in cuprate superconductors emerges in close proximity to the long-range antiferromagnetic Mott insulator with a spin density wave (SDW) order \cite{R14,R15}, it is belived that a magnetic QCP, separating SC and the coexistence of SC $+$ SDW orders, exists on the underdoped regime beneath the superconducting dome as a result of the quantum interplay of SC and SDW \cite{R16,R17,R18}. Actually, the unambiguous experimental evidence of the coexistence of SDW and SC was observed at the strongly underdoped regime from numerous NMR and neutron scattering experiments on cuprate superconductors \cite{R19,R20,R21,R22,R23,R24}. As shown in Fig. 2, the fact that $\Delta T/T_c$ is zero at $p$ $\simeq$ 0.075 means that the phase correlation length between the superconducting grains diverges as $\mid p-p_{c1}\mid$ vanishes ($p_{c1}$=0.075), which indicates the presence of QCP at $p$ $\simeq$ 0.075. Vishik $et$ $al.$ \cite{R3} reported that  in Bi-2212, the gap slope, which represents the degree of the increase of the $d$-wave gap as a function of the momentum away from the node of the Fermi surface, changed suddenly at $p$ $\simeq$ 0.075, at which $\Delta T/T_c$ is zero, as shown in Fig. 2. The SDW
volume fraction at $T$ = 0 K, as measured by muon spin rotation, disappears in the vicinity of the magnetic QCP ($p_{c1}$ $\simeq$ 0.075)\cite{R25}. The strong coupling of the electrons to spin fluctuations at the magnetic QCP is likely to enhance the coherent motion of the Cooper pairs between the different superconducting grains. 

Moon and Sachdev \cite{R26} predicted the presence of a crossover line between the SC $+$ SDW and SC orders, which begins at the magnetic QCP ($p_{c1}$) at $T$ = 0 K and ends at $T_c$ of a $p_m$ ($>$ $p_{c1}$) of the superconducting dome. Therefore, as $p$ approaches from $p_{c1}$ to $p_m$, the coupling of the electron to spin fluctuation weakens, resulting in an increase in $\Delta T/T_c$. Based on these experimental results, it is argued that the crossover line ends at $p$ $\simeq$ 0.10, above which $\Delta T/T_c$ remains at an almost constant value of 0.140. The data of $\Delta T/T_c$ near $p$ = 0.075 is described well by a solid line in Fig. 2, given by the function of $A$tanh($\mid p-p_{c1}\mid$/$c$), where $A$ and $c$ are 0.0140 and 0.014, respectively. As tanh$x$ $\approx$ $x$ for $x$ $\ll$ 1, $\Delta T/T_c$ $\approx$ 10$\mid p-p_{c1}\mid$ near $p_{c1}$.    

It was suggested theoreticallythat a magnetic QCP triggers the emergence of orders with different symmetries from those of the parent transition \cite{R27,R28}. These orders include $d$-wave SC, charge density wave (CDW) with a checkerboard structure, and a pseudogap (PG) state lacking long range order. In the La-based cuprate superconductor, the stripe-like charge order (CO) near $p$ = 1/8 is associated with the anomalous suppression of $T_c$. Recently, Chang $et$ $al.$ \cite{R29} and Ghiringhelli $et$ $al.$ \cite{R30} reported the observations of an incipient CDW, which is a periodic modulation of the conduction electron density, by high-energy X-ray diffraction in underdoped YBCO. The peak intensity of CDW emerged below $T_{CDW}$ $\sim$ 135 K, which is lower than $T^*$ $\sim$ 220 K at the same doping level, where $T^*$ is the temperature, at which PG opens. The peak intensity grows upon cooling to $T_c$ below, which it is partially suppressed. This suggests that CDW and SC are competing orders \cite{R31}.

Comin $et$ $al.$ \cite{R32} and da Silva Neto $et$ $al.$ \cite{R33} reported that the emergence of CDW order, which was observed previously in LSCO and YBCO, is a ubiquitous phenomenon in other cuprate superconductors, such as Bi$_2$Sr$_2$CuO$_{6+\delta}$ (Bi-2201) and Bi-2212 from scanning tunneling microscopy, ARPES and bulk resonant elastic X-ray scattering experiments. They also reoprted the competing nature of SC and CDW by observing that the onset of SC weakens the strength of CO. In particular, da Silva Neto $et$ $al.$ \cite{R33} found that from $p$ = 0.12 to the underdoped regime in Bi-2212, the CO wave vector, $\delta$, increases from 0.25 (nearly commensurate) to 0.30 (incommensurate) in the narrow doping range ($\Delta$$p$ $\sim$ 5.0$\times$10$^{-3}$) near $p$ = 0.1, which is in contrast to the expectation from a Fermi surface nesting mechanism. Such an increase of $\delta$ in the narrow doping range near $p$ = 0.1 is in line with the steep suppression of $\Delta T/T_c$ near $p$ $\simeq$ 0.1 in Fig. 2.

As $p$ goes from $p$ = 0.1 to the overdoped regime, both $\Delta T/T_c$ and $T_c$/$\rho_s(0)$ move into suppression at $p$ $\simeq$ 0.15. In contrast $T_c$/$\rho_s(0)$ maintains a tendency to suppress with increasing $p$, but $\Delta T/T_c$ increases at $p$ $\simeq$ 0.19, as shown in Fig. 2. As suggested above, the $p$ dependence of $\Delta T/T_c$ in 0.10 $<$ $p$ $<$ 0.19 is dominated by the same mechanism, $i$.$e$. the coexistence of CDW, PG, and SC. The associated suppression in $\Delta T/T_c$ and $T_c$/$\rho_s(0)$ at $p$ $\simeq$ 0.15 can be  explained naturally by the competing nature of CDW and PG on SC. The competition between orders implies that the order, which is set up first, tends to suppress the other on the Fermi surface \cite{R27,R28}. Although both $T_{CDW}$ and $T^*$ decrease with increasing $p$, our experimental results suggest that the two ($T_{CDW}$ and $T^*$) are larger than $T_c$ up to $p$ $\simeq$ 0.15. Therefore the magnitudes of $\Delta T/T_c$ and $T_c$/$\rho_s(0)$ remain almost unchanged. In contrast, for $p$ $>$ 0.15, $T_{CDW}$, which is smaller than $T^*$, first becomes smaller than $T_c$, and then, $T^*$ becomes smaller than $T_c$ at a slightly higher $p$, giving rise to an enhancement in $\rho_s(0)$. These considerations on $T_{CDW}$ and $T^*$ are consistent with the phase diagram given in Ref. 34. Despite the decrease in $T_c$ for $p$ $>$ 0.16, such enhancement in $\rho_s(0)$ keeps until both $T_{CDW}$ and $T^*$ vanish, which is responsible for the decrease in $\Delta T/T_c$ at $p$ $>$ 0.15. 

As $p$ increases, the CDW and PG orders vanish simultaneously at $T$ = 0 K at a special $p$ in the superconducting dome, setting up a charge QCP, $p_{c2}$, at some distance from the magnetic QCP \cite{R27}. Recently, Fujita $et$ $al.$ \cite{R35} reported that the electronic symmetry breaking tendencies, closely related to CDW and PG, weaken with increasing $p$ and disappear close to $p$ $\simeq$ 0.19. Therefore, $\rho_s(0)$ reaches its maximum due to the disappearance of competing orders, CDW and PG, on SC at $p$ $\simeq$ 0.19. As a result, $\Delta T/T_c$ approaches the minimum value, not zero, at $p$ $\simeq$ 0.19, because both the CDW and PG orders lack a long range correlation length, which is in contrast to what is observed at $p$ $\simeq$ 0.075. The order for $p$ $>$ 0.19 in the superconducting dome is a pure $d$-wave SC, which results in the increase in $\Delta T/T_c$ due to the enhanced phase fluctuations by the decrease in $\rho_s(0)$ with further increasing $p$ from $p$ $\simeq$ 0.19. A previous study reported that $H_{c2}(0)$ of Bi-2212 with $p$ $>$ 0.16 is a maximum at $p$ $\simeq$ 0.19 \cite{R5}, similar to YBCO mentioned above. The Fermi surface, on which coherent Bogoliubov quasiparticles are detected, undergoes an abrupt transition from arcs to a closed contour at $p$ $\simeq$ 0.19 \cite{R35,R36}. This suggests that in addition to the magnetic QCP at $p_{c1}$ $\simeq$ 0.075, there is another QCP, $i.e.$ a charge QCP, at $p_{c2}$ $\simeq$ 0.19 at $T$ = 0 K beneath the superconducting dome.        

In summary, the phase coherences between the Bi-2212 superconducting grains with $p_{c1}$ $\simeq$ 0.075 and $p_{c2}$ $\simeq$ 0.190 are enhanced remarkably. In particular, our experimental results at the strongly underdoped regime deviate from expectation based on the doping level dependence of the superfluid density at $T$ = 0 K. These findings are governed by quantum interplay between the competing orders, SDW, CDW, PG, and SC, inside the superconducting dome of cuprate superconductors. Two quantum critical points at $p_{c1}$ $\simeq$ 0.075 and $p_{c2}$ $\simeq$ 0.190 are likely to exist beneath the superconducting dome.

\begin{acknowledgements}
 This study was financially supported by the Research Fund Program of Research Institute for Basic Science, Pusan National University, Korea, 2013, Project No. RIBS-PNU-2013-111.
\end{acknowledgements}

\newpage

{\bf Figure Captions}\\

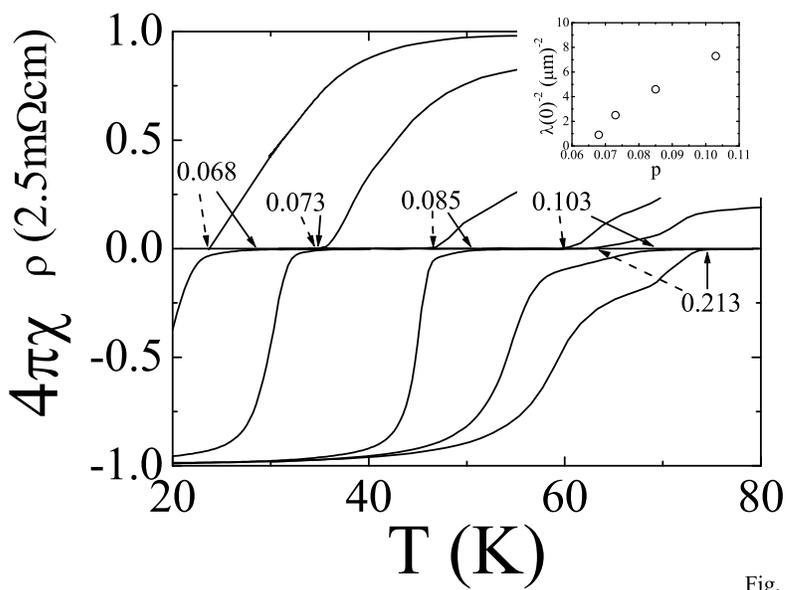
Figure 1. Temperature dependence of the ac susceptibility and resistivity for Bi-2212 with different $p$ values. The solid arrow and  dashed one indicate $T_c$ and $T_{\phi}$ for each sample, respectively. The inset shows the $p$ dependence of $\lambda(0)^{-2}$ as the Mott insulating state is approached.\\

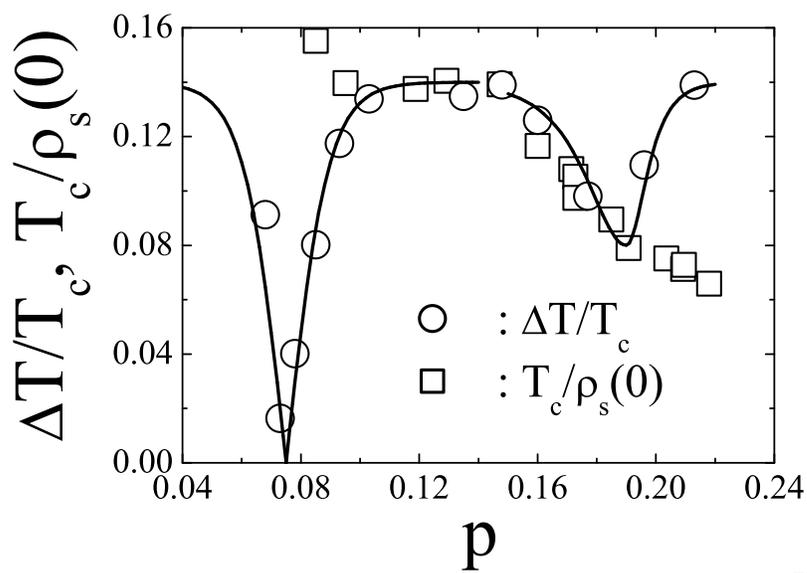
Figure 2. $p$ dependence of $\Delta T/T_c$ ( $ \Delta T$ $\equiv$ $T_c-T_{\phi}$) and $T_c$/$\rho_s(0)$ for Bi-2212. The solid line near $p$ = 0.075 shows a fit by a function of $A$tanh($\mid p-p_{c1}\mid$/$c$), where $A$ and $c$ are 0.0140 and 0.014, and the solid line near $p$ = 0.19 is a guide for the eye.\\

\newpage

\end{document}